\documentclass[iop]{emulateapj}

\usepackage{multirow}

\newcommand{\xmm} {{\it XMM-Newton}}
\newcommand{\chandra} {{\it Chandra}}
\newcommand{\nustar} {{\it NuSTAR}}

\newcommand{\swiftbat} {{\it Swift}/BAT}
\newcommand{\hst} {{\it Hubble Space Telescope}}

\newcommand{\cmsq} {cm$^{-2}$}
\newcommand{\nh} {$N_{\rm{H}}$}
\newcommand{\lx} {$L_{\rm{X}}$}

\newcommand{\degree}{{$^\circ$}}

\newcommand{\ergs}{\mbox{\thinspace erg\thinspace s$^{-1}$}}

\newcommand{\lbol} {$L_{\rm Bol}$}
\newcommand{\kbol} {$\kappa_{\rm Bol}$}
\newcommand{\mbh} {$M_{\rm BH}$}

\newcommand{\lamedd} {$\lambda_{\rm Edd}$}

\newcommand{\thetator} {$\theta_{\rm tor}$}
\newcommand{\msol} {$M_{\odot}$}
\newcommand{\lsol} {$L_{\odot}$}

\shorttitle{X-ray bolometric corrections for CT AGN}
\shortauthors{Brightman et al.}

\begin{document}

\title{X-ray bolometric corrections for Compton-thick active galactic nuclei}

\author{M. Brightman$^{1}$, M. Balokovi\'{c}$^{1}$, D. R. Ballantyne$^{2}$, F. E. Bauer$^{3,4,5}$, P. Boorman$^{6}$, J. Buchner$^{3}$, W. N. Brandt$^{7,8,9}$, A. Comastri$^{10}$, A. Del Moro$^{11}$, D. Farrah$^{12}$ P. Gandhi$^{6}$, F. A. Harrison$^{1}$, M. Koss$^{13}$, L. Lanz$^{14}$, A. Masini$^{10,15}$, C. Ricci$^{3,16}$, D. Stern$^{17}$, R. Vasudevan$^{18}$ D. J. Walton$^{18}$}

\affil{$^{1}$Cahill Center for Astrophysics, California Institute of Technology, 1216 East California Boulevard, Pasadena, CA 91125, USA\\
$^{2}$Center for Relativistic Astrophysics, School of Physics, Georgia Institute of Technology, Atlanta, GA 30332, USA\\
$^{3}$Instituto de Astrof\'{i}sica, Facultad de F\'{i}sica, Pontificia Universidad Catolica de Chile, Casilla 306, Santiago 22, Chile\\
$^{4}$Millennium Institute of Astrophysics\\
$^{5}$Space Science Institute, 4750 Walnut Street, Suite 205, Boulder, CO 80301, USA\\
$^{6}$School of Physics \& Astronomy, University of Southampton, Highfield, Southampton SO17 1BJ, UK\\
$^{7}$Department of Astronomy and Astrophysics, 525 Davey Lab, The Pennsylvania State University, University Park, PA 16802, USA\\
$^{8}$Institute for Gravitation and the Cosmos, The Pennsylvania State University, University Park, PA 16802, USA\\
$^{9}$Department of Physics, 104 Davey Laboratory, The Pennsylvania State University, University Park, PA 16802, USA\\
$^{10}$INAF Osservatorio Astronomico di Bologna, via Gobetti 93/3, I-40129 Bologna, Italy \\
$^{11}$Max-Planck-Institut f\"{u}r extraterrestrische Physik, Giessenbachstrasse 1, D-85748, Garching bei M\"{u}nchen, Germany\\
$^{12}$Department of Physics, Virginia Tech, Blacksburg, VA 24061, USA\\
$^{13}$SNSF Ambizione Fellow, Institute for Astronomy, Department of Physics, ETH Zurich, Wolfgang-Pauli-Strasse 27, CH-8093 Zurich, Switzerland\\
$^{14}$Department of Physics and Astronomy, Dartmouth College, 6127 Wilder Laboratory, Hanover, NH 03755, USA\\
$^{15}$Dipartimento di Fisica e Astronomia (DIFA), Universit\'{a} di Bologna, viale Berti Pichat 6/2, 40127 Bologna, Italy\\
$^{16}$Kavli Institute for Astronomy and Astrophysics, Peking University, Beijing 100871, China\\
$^{17}$Jet Propulsion Laboratory, California Institute of Technology, Pasadena, CA 91109, USA\\
$^{18}$Institute of Astronomy, Madingley Road, Cambridge CB3 0HA, UK\\}

\begin{abstract}

We present X-ray bolometric correction factors, \kbol\ ($\equiv$\lbol/\lx), for Compton-thick (CT) active galactic nuclei (AGN) with the aim of testing AGN torus models, probing orientation effects, and estimating the bolometric output of the most obscured AGN. We adopt bolometric luminosities, \lbol, from literature infrared (IR) torus modeling and compile published intrinsic 2--10 keV X-ray luminosities, \lx, from X-ray torus modeling of \nustar\ data. Our sample consists of 10 local CT AGN where both of these estimates are available. We test for systematic differences in \kbol\ values produced when using two widely used IR torus models and two widely used X-ray torus models, finding consistency within the uncertainties. We find that the mean \kbol\ of our sample in the range \lbol$\approx10^{42}-10^{45}$ \ergs\ is log$_{10}$\kbol$=1.44\pm0.12$ with an intrinsic scatter of $\sim0.2$ dex, and that our derived \kbol\ values are consistent with previously established relationships between \kbol\ and \lbol\ and \kbol\ and Eddington ratio (\lamedd). We investigate if \kbol\ is dependent on \nh\ by comparing our results on CT AGN to published results on less-obscured AGN, finding no significant dependence. Since many of our sample are megamaser AGN, known to be viewed edge-on, and furthermore under the assumptions of AGN unification whereby unobscured AGN are viewed face-on, our result implies that the X-ray emitting corona is not strongly anisotropic. Finally, we present \kbol\ values for CT AGN identified in X-ray surveys as a function of their observed \lx, where an estimate of their intrinsic \lx\ is not available, and redshift, useful for estimating the bolometric output of the most obscured AGN across cosmic time.

\end{abstract}


\section{Introduction}

The bolometric luminosity, \lbol, of an accreting supermassive black hole (SMBH), otherwise known as an active galactic nucleus (AGN), describes the integrated emission from the accretion process, which traces the mass accretion rate onto the SMBH ($L_{\rm Bol}=\eta\dot{m}c^{2}$, where $\dot{m}$ is the mass accretion rate and $\eta$ the accretion efficiency). Thus \lbol\ is an important parameter for understanding the growth of SMBHs. The emission from the accretion disk, which is the primary power generation mechanism, is reprocessed by a number of components in the vicinity of the disk, one of which is a hot corona of electrons that Compton scatters the optical and UV disk emission into the X-ray regime \citep[e.g.][]{haardt91,haardt93}.

The fraction of the disk emission that is up-scattered in to the X-ray regime is parameterized by the X-ray bolometric correction factor (from here on \kbol), which is defined as \lbol/\lx, where \lx\ is the X-ray luminosity in the 2--10 keV band. Many works have investigated \kbol, finding that it is dependent on \lbol\ \citep[e.g.][]{marconi04,steffen06,hopkins07,lusso12,liu16} and Eddington ratio \citep[\lamedd$\equiv$\lbol/$L_{\rm Edd}$, where $L_{\rm Edd}=4\pi$G\mbh\ $m_{\rm p}c/\sigma_{\rm T}\simeq1.26\times10^{38}$(\mbh/\msol) \ergs\ and \mbh\ is the mass of the black hole, e.g.][]{wang04,vasudevan07,vasudevan09,lusso10,lusso12,jin12,fanali13,liu16}. 

Characterizing \kbol\ and its dependencies is important for understanding accretion physics and for estimating \lbol\ when it is not possible to observe the intrinsic disk emission, but where \lx\ is known. This can be the case for obscured AGN, where gas and dust in the line of sight extinguishes the optical and UV emission from the accretion disk but X-rays from the corona penetrate through (for all but the most extreme absorbing columns \nh$<10^{24}$ \cmsq). While the dependencies of \kbol\ have been well established for unobscured, type 1 AGN, only a few studies have focussed on obscured, type 2 AGN \citep[e.g.][]{pozzi07,vasudevan10,lusso11,lusso12}. 

Investigating \kbol\ for obscured AGN is important since the majority of AGN in the Universe are obscured \citep[e.g.][]{martinezs05,ueda14,buchner15,aird15a}. It also has potential for testing the AGN unification scheme \citep[e.g.][]{antonucci93,urry95}, the simplest form of which describes the differences between type 1 and type 2 AGN as solely due to orientation, where type 2 AGN are more inclined systems and our view of the central engine is through a toroidal structure of gas and dust. The most extremely obscured sources, so-called Compton-thick (CT) AGN (\nh$>1.5\times10^{24}$ \cmsq) constitute some $\sim20-40$\% of the AGN population \cite[e.g.][]{burlon11,brightman11,brightman12b,buchner15} and host some of the most highly inclined systems, revealed through the detection of disk megamasers \citep{zhang06,masini16}. However, for CT AGN, flux suppression is high even in the X-ray band and the effect of Compton scattering on the X-ray spectrum is dependent on the geometry of the obscuring material \citep[e.g.][]{brightman15} making the intrinsic \lx\ difficult to estimate. For this reason \kbol\ has not previously been investigated for CT AGN.

\begin{figure}
\begin{center}
\includegraphics[width=90mm]{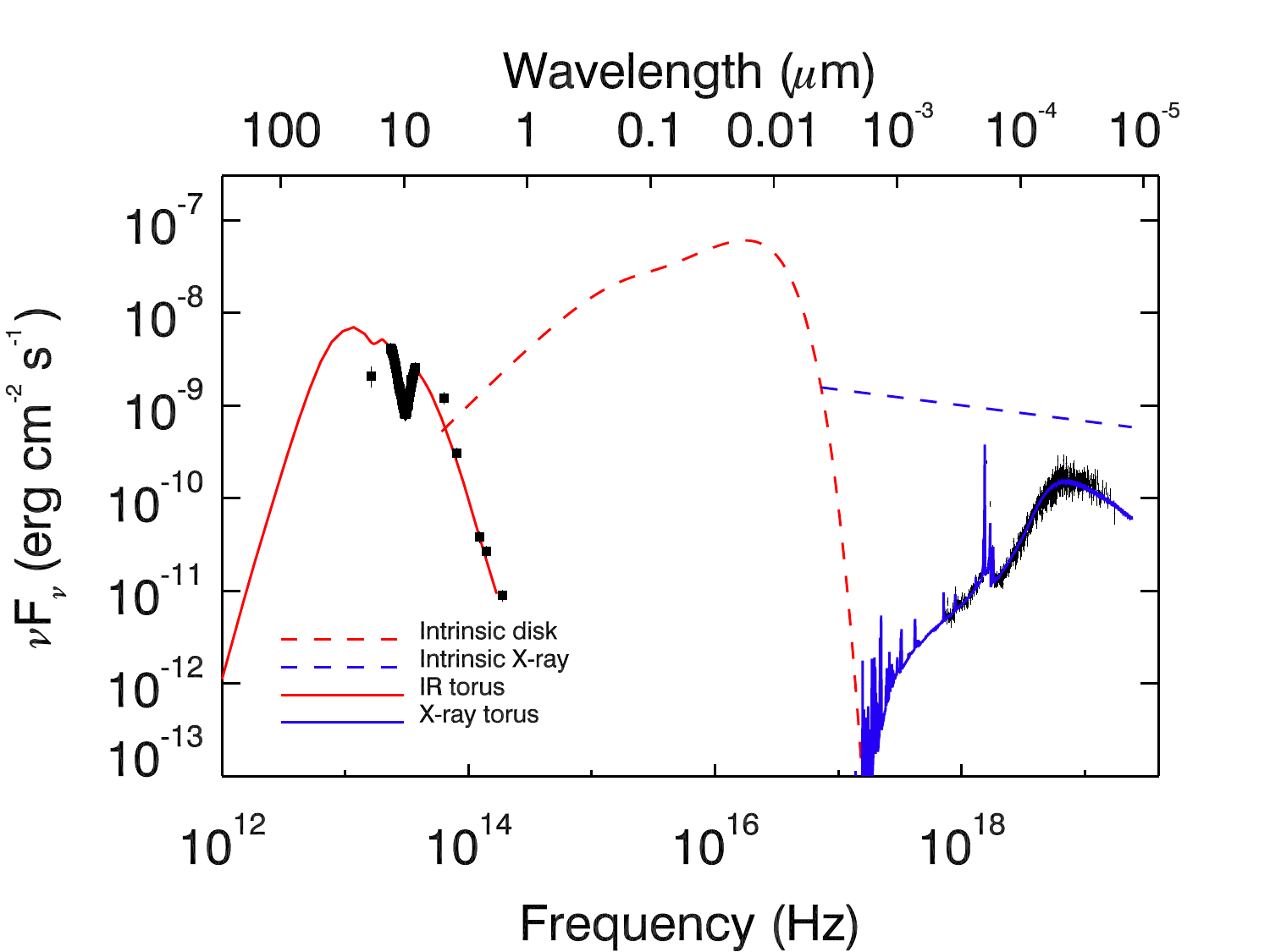}
\caption{IR to X-ray SED of the CT AGN in the Circinus galaxy. The IR data points consist of high-spatial resolution 8--13 $\mu$m spectra from the Thermal-Region Camera Spectrograph (TReCS) on Gemini-South, described in \cite{roche06}, and high-spatial resolution NIR and MIR photometry from ground-based observations and \hst/NICMOS observations described in \cite{alonso11}. The solid red line is a fit to these data with the {\sc clumpy} IR torus model of \cite{nenkova08} by \cite{ichikawa15}, which yielded the \lbol\ estimate. The dashed red line represents the inferred intrinsic accretion disk emission, given the known black hole mass and the inferred \lbol\ from the {\tt optxagn} model \citep{done12}. The X-ray data are from \nustar, described in \cite{arevalo14}, fitted with the X-ray {\tt torus} model by \cite{brightman15}, plotted as a solid blue line. The intrinsic X-ray spectrum inferred from this model is plotted as a dashed blue line, from which we obtain our intrinsic \lx\ estimate. The gap between the dashed and solid blue lines is due to absorption.}
\label{fig_sed}
\end{center}
\end{figure}

At energies $>$10 keV, while the effect of Compton scattering remains, the flux suppression is lower due to the declining photoelectric absorption cross section with increasing energy. Therefore, \nustar\ \citep{harrison13}, with its sensitivity at these energies, is ideal for estimating the intrinsic \lx\ for CT AGN. For this, X-ray spectral models that take into account the absorption and Compton scattering are needed \citep[e.g.][]{ikeda09,murphy09,brightman11,liu14}. Figure \ref{fig_sed} illustrates this point, showing the \nustar\ data of the well-known CT AGN in the Circinus galaxy \citep{arevalo14}, fitted with the \cite{brightman11} {\tt torus} model, also showing the intrinsic X-ray spectrum inferred using the model parameters. The figure shows that a greater fraction of X-ray flux emerges above 10 keV in the source, than below 10 keV. Since its launch in 2012, \nustar\ has observed a large number of CT AGN, with \lx\ estimated from both the {\tt mytorus} model of \cite{murphy09} and the {\tt torus} model of \cite{brightman11} by various authors \citep[e.g.][]{puccetti14,arevalo14,balokovic14,gandhi14,bauer15,brightman15,koss15,annuar15,rivers15b,marinucci16,ricci16,masini16,farrah16,boorman16}.

As well as being reprocessed by the hot corona into the X-rays, the AGN disk emission is also reprocessed by the dust in the torus into the infrared \citep[e.g.][]{pier92}. The structure of the dust torus does not necessarily have the same geometry as the X-ray absorbing material, which is gas that can exist within the dust sublimation radius. As in the X-ray band, torus models have been calculated to model the infrared emission \citep[e.g.][]{nenkova08,hoenig10b,stalevski12,efstathiou13}. A natural parameter derived from these models is \lbol. Since significant infrared emission is also emitted by dusty star formation in the host galaxy, high-spatial resolution IR data or broadband spectral energy distribution (SED) modeling are required to isolate the AGN and model the torus emission \citep[e.g.][]{farrah03,stierwalt14}. \cite{alonso11} presented the results from fitting of \cite{nenkova08} {\sc clumpy} torus model to high-spatial resolution IR spectroscopy and photometry of 13 nearby Seyfert galaxies, finding that their \lbol\ estimates agreed well with other estimates from the literature. A further expanded study in the IR was conducted by \cite{ichikawa15}, which presented an analysis of 21 nearby AGN, with significant overlap with the sample of AGN with X-ray torus modeling.

One such source in common is the CT AGN in the Circinus galaxy. Along with the \nustar\ data in Figure \ref{fig_sed}, we plot the high spatial resolution IR data along with the fit using the IR torus model. The inferred accretion disk spectrum is also shown.

The aim of this paper is to take advantage of the recent advances in both IR and X-ray torus modeling that produce estimates of \lbol\ and intrinsic \lx\ respectively and derive \kbol\ values for CT AGN. We start in Section \ref{sec_sample} where we describe our sample selection. In Section \ref{sec_ctkbol} we collect and compare results from the literature on the two widely used X-ray torus models, {\tt mytorus} \citep{murphy09} and {\tt torus} \citep{brightman11} and two widely used IR torus models from \cite{fritz06} and \cite{nenkova08}. We assess the systematic differences, if any. Following this we test if the \kbol\ values we estimate for CT AGN are consistent with established relationships between \kbol\ and  \lbol\ and \kbol\ and \lamedd\ as determined from unobscured AGN. Next we compare our new \kbol\ results for CT AGN to results from previous studies for less obscured systems in order to explore any dependence of \kbol\ on \nh\ and probe orientation effects. We then present \kbol\ for CT AGN as a function of observed \lx\ and redshift, useful for studies of CT AGN in surveys where there is not a good estimate of the intrinsic \lx. We discuss our results in Section \ref{sec_disc} and present our conclusions in Section \ref{sec_conc}. We define \lbol\ as the total of the inferred disk emission (from IR torus modeling) together with the intrinsic \lx\ (from X-ray torus modeling) in order to be consistent with previous works \citep[e.g.][]{marconi04,vasudevan10}. We assume a flat cosmological model with $H_{\rm 0}$=70 km s$^{-1}$ Mpc$^{-1}$ and $\Omega_{\Lambda}$=0.73.

\section{Sample Selection and Luminosity Estimates}
\label{sec_sample}

We compile \lx\ measurements from X-ray torus modeling of \nustar\ data and \lbol\ results from IR spectral/SED modeling from the literature, finding 10 local CT AGN where both of these exist. We find five sources from the sample of \cite{ichikawa15}, who used the {\sc clumpy} torus models of \cite{nenkova08} to calculate \lbol, fitting over the range 1.25--30 $\mu$m. We find a further four sources from the sample of \cite{gruppioni16}, who rather than using high-spatial resolution IR data to isolate the AGN emission, carry out SED decomposition to isolate the AGN emission from the host galaxy, using the approach described by \cite{berta13}. They use the torus model of \cite{fritz06}, which models a smooth distribution of dust and calculate \lbol\ over the 1--1000 $\mu$m range.  Finally, \cite{woo02} calculated \lbol\ for a large number of AGN by simply integrating over the observed multiwavelength SED. This was a far less sophisticated approach to \lbol\ estimation than IR torus modeling since it presumably does not account for host-galaxy emission. We compare these \lbol\ estimates for four sources where overlap with the IR torus modeling exists. We find one CT AGN where X-ray torus modeling has been conducted and an \lbol\ estimate exists from \cite{woo02}, NGC~2273, which we include in our sample. 

We list some basic observational properties of our sample in Table \ref{tab_sample}. Due to the detailed torus modeling involved, these sources are necessarily nearby ($D<60$ Mpc). Our sample also contains six megamaser AGN indicating that they have high inclinations, since these are required to produce this emission. Furthermore, the Keplerian motion of the masing material provides an accurate measurement of \mbh\ \citep[e.g.][]{kuo11} and allows us to test the relationship between \kbol\ and \lamedd. We also list the \mbh\ estimates in Table \ref{tab_sample}. 

The different torus models used to calculate \lx\ and \lbol\ have properties that are inherent to each, which we describe here.  The \cite{nenkova08} models assume a dust torus consisting of clouds that are distributed with axial symmetry and whose number per unit length depends on the distance from the illuminating source and the angle from the equatorial plane. This torus is illuminated by an intrinsic disk spectrum which takes the form of a piecewise power-law distribution described in \cite{rrobinson95}, where $\lambda F_{\lambda}=\lambda^{1.2}$ for $\lambda\leq0.01 \mu$m, $\lambda F_{\lambda}=$ constant for $0.01\leq\lambda<0.1 \mu$m, $\lambda F_{\lambda}=\lambda^{-0.5}$ for $0.1\leq\lambda<1 \mu$m and $\lambda F_{\lambda}=\lambda^{-3}$ for $1 \mu$m$\leq\lambda$. Integrating over this assumed disk spectrum yields \lbol. The anisotropy of this clumpy torus is discussed at length in \cite{nenkova08} and depends on the various parameters of the torus. For example, the torus becomes less anisotropic when the power-law index of the radial distribution of clouds increases, i.e. steeper. This is a free parameter in the model and hence fitted for in SED modeling. The anisotropy is also strongly wavelength dependent, with the torus being being particularly isotropic at 12$\mu$m.

While the \cite{nenkova08} model assumes a clumpy distribution of dust, the \cite{fritz06} model also assumes smooth distribution, but that also depends on the radial distance from the source and the equatorial angle. An intrinsic disk spectrum that illuminates the torus isotropically in the form of a piecewise power-law distribution that is similar but not identical to that assumed by the \cite{nenkova08} models. Here $\lambda F_{\lambda}\propto\lambda^{1.2}$ for $0.001\leq\lambda\leq0.03 \mu$m, $\lambda F_{\lambda}\propto$ constant for $0.03\leq\lambda<0.125 \mu$m and $\lambda F_{\lambda}\propto\lambda^{-0.5}$ for $0.125\leq\lambda<20 \mu$m. The degree of anisotropy from this torus is rather higher than for the clumpy torus, and depends on the viewing angle and the equatorial optical depth. Again these are free parameters of the model and are fitted for in SED modeling. \lbol\ is calculated from a bolometric correction factor given the best-fit template \cite{gruppioni16}.

The X-ray torus models of \cite{murphy09} and \cite{brightman11} both model smooth distributions of gas. {\tt mytorus} assumes a `doughnut'-like geometry with a circular cross-section, whereas the {\tt torus} model assumes a `spherical' torus with a biconical cut out. Both models assume a intrinsic source spectrum that takes power-law form with $F_{\gamma}\propto E^{-\Gamma}$ ($\lambda F_{\lambda}=E^{-\Gamma+2}$). For sight lines through the torus, the anisotropy in the \nustar\ band is negligible.

The luminosities that we have compiled here are a collection of literature values that also depend on the distance assumed by each author, which can often have large discrepancies due to the nearby nature of these galaxies. For example, \cite{brightman15} assume a distance of 6.2~Mpc to the Circinus galaxy based on the Hubble flow distance for the intrinsic \lx\ estimate from the {\tt torus} model, whereas \cite{arevalo14} assume a distance of 4.2~Mpc based on the Tully estimate for the intrinsic \lx\ estimate from the {\tt mytorus} model. Furthermore, \cite{ichikawa15} assume a distance of 4~Mpc for the \lbol\ estimate. Since luminosity scales with distance squared, this difference leads to a factor of $\sim2$ discrepancy which we must account for when calculating and comparing \kbol\ values. We do this by taking the luminosity and the distance assumed by each author and correcting the luminosity assuming the distance that we list in Table \ref{tab_sample}. 

We list the intrinsic \lx\ and \lbol\ estimates in Table \ref{tab_lum} along with the corresponding \kbol\ values which have been corrected for distance. Our sample spans a range of \lx$\approx10^{41.5}-10^{44}$ \ergs, \lbol$\approx10^{42}-10^{45}$ \ergs, \mbh$\approx10^{6}-7\times10^{7}$ \msol\ and \lamedd$\approx0.01-0.3$. All our sources are Compton thick by selection with \nh$=10^{24}-10^{25}$ \cmsq, with the exception of NGC~1320 that has \nh$>10^{25}$ \cmsq\ \citep{brightman15}.

\begin{table}
\centering
\caption{Basic properties of the galaxies in our sample}
\label{tab_sample}
\begin{center}
\begin{tabular}{l c c c c c c c c c c c}
\hline
Name 	& Mag	& Morphology 	& Distance		& \mbh	& Ref  \\
(1) & (2) & (3) & (4) & (5) & (6)  \\
\hline
Circinus 		& 10.0 ($I$)		&	SA(s)b?		&	4.2		& 1.7$\pm$0.3 & a 	\\
NGC 424 		& 12.8 ($I$)		&	(R)SB0/a?(r)	&	50.6		&	\\
NGC 1068 	& 9.9 ($I$)		&	(R)SA(rs)b	&	14.4		& 8.0$\pm$0.3	& b \\
NGC 1194		& 12.5 ($i$)		& 	SA0\^\ +?		&	58.9		& 65.0$\pm$3.0& c\\
NGC 1320 	& 12.5 ($V$)	&	Sa? edge-on	&	39.1		& \\
NGC 1386 	& 10.76 ($R$)	&	SB0\^\ +(s)	&	15.9		& 1.2$\pm$1.1	& d\\
NGC 2273	& 10.15 ($I$)		&	SB(r)a?		&	28.9		& 7.5$\pm$0.4	& c\\
NGC 3079 	& 9.5 ($I$)		&	SB(s)c edge-on	&	19.2		& 2.4$^{+2.4}_{-1.2}$ & e\\
NGC 5643	& 10.6 ($I$)		&	SAB(rs)c		& 	13.9		&	\\
NGC 7582	& 9.2 ($I$)		&	(R$^{\prime}$)SB(s)ab	&	22		&	\\

\hline
\end{tabular}
\tablecomments{Column (1) lists the galaxy name, Column (2) gives the visual magnitude (band in parentheses (Cousins $I$-band where available), Column (3) shows the galaxy morphology classification from NED, Coulmn (4) gives the assumed distance to the source in Mpc, and Column (5) presents \mbh\ in units of $10^6$ \msol\ where this has been estimated from the megamaser emission with the reference for this given in Column (6). References: a. \cite{greenhill03}, b. \cite{lodato03}, c. \cite{kuo11}, d. \cite{mcconnell13}, and e. \cite{kondratko05}. }

\end{center}
\end{table}

\begin{table*}
\centering
\caption{X-ray and bolometric luminosities of the sample}
\label{tab_lum}
\begin{center}
\begin{tabular}{l c c c c c c c c c c c c c}
\hline
Name 	& \lx\ 			& Ref 	& \lx\ 			& Ref 	& \lbol\ 		& Ref	& \lbol\ 			& Ref	& \kbol	& \kbol	 \\
			& ({\tt torus})	& 		& ({\tt mytorus})	&		& (IR torus)	&		& (SED integration)	& 	& ({\tt torus})	& ({\tt mytorus})	 \\	
(1) & (2) & (3) & (4) & (5) & (6) & (7) & (8) & (9) & (10) & (11) \\
\hline
Circinus 		&	42.51$^{+	0.07	}_{-0.09}$ &	a	& 42.58	&	e	& 43.5$\pm$0.6	& i	& 43.60	& j	& 1.39$\pm$0.61 	& 1.01$\pm$0.67 \\
NGC 424 		&	43.96$^{+	0.21	}_{-0.16}$ &	a	& 43.50	&	f	& 44.77$\pm$0.01	& h	&		&	& 0.88$\pm$0.19	& 1.30$\pm$0.30 \\
NGC 1068 	&	42.87$^{+	0.04	}_{-0.06}$ &	a	& 43.34	&	g	& 44.4$\pm$0.5	& i	& 44.98	& k	& 1.61$\pm$0.50	& 1.06$\pm$0.58 \\
NGC 1194		&	42.56				&	b	& 42.75	&	b	& 44.74$\pm$0.04	& h	&		&	& 2.19$\pm$0.30	& 1.98$\pm$0.30 \\
NGC 1320 	&	42.79$^{+	0.12	}_{-0.09}$ &	a	& 42.85	&	f	& 44.16$\pm$0.06	& h	& 44.02	& k	& 1.40$\pm$0.12	& 1.36$\pm$0.31\\
NGC 1386 	&	41.84$^{+	0.26	}_{-0.05}$ &	a	& 41.30	&	b	& 42.5$\pm$0.5	& i	& 43.38	& k	& 0.83$\pm$0.52	& 1.33$\pm$0.58\\
NGC 2273	&	43.11$^{+	0.19	}_{-0.34}$ &	b	& 42.60	&	b	&				& 	& 44.05	& k	& 1.04$\pm$0.40 	& 1.52$\pm$0.42\\
NGC 3079 	&	41.53$^{+	0.45	}_{-0.43}$ &	a	& 42.15	&	b	& 43.61$\pm$0.25	& h	&		&	& 2.09$\pm$0.51	& 1.49$\pm$0.39\\
NGC 5643	&	41.90				&	c	& 41.95	&	c	& 43.00$\pm$0.5	& i	&		&	& 1.02$\pm$0.58	& 0.98$\pm$0.58\\
NGC 7582	&	41.70				&	d	& 41.54	&	d	& 43.50$\pm$0.5	& i	&		&	& 1.85$\pm$0.58	& 2.00$\pm$0.58\\

\hline
\end{tabular}
\tablecomments{Column (1) lists the AGN name, Column (2) presents the logarithm of the intrinsic 2--10 keV luminosity in \ergs\ estimated from the {\tt torus} model of \cite{brightman11}, with references listed in Column (3). Column (4) presents the logarithm of the intrinsic 2--10 keV luminosity in \ergs\ estimated from the {\tt mytorus} model of \cite{murphy09}, with references listed in Column (5). Column (6) lists the logarithm of the bolometric luminosity in \ergs\ estimated from IR torus modeling, with references in Column (7). Column (8) list the logarithm of the bolometric luminosity in \ergs\ estimated from SED integration with references in Column (9). Column (10) lists the logarithm of \kbol\ when using the \lx\ measurement from the {\tt torus} model and \lbol\ from the IR torus modeling, corrected for distance discrepancies between the two, and Column (11) lists the logarithm of \kbol\ when using the \lx\ measurement from the {\tt mytorus} model, also using \lbol\ from the IR torus modeling, corrected for distance discrepancies. References: a. \cite{brightman15}, b. \cite{masini16}/private communication, c. \cite{annuar15}, d. \cite{rivers15b}, e. \cite{arevalo14}, f. \cite{balokovic14}, g. \cite{bauer15},  h. \cite{gruppioni16}, i. \cite{ichikawa15}, j. \cite{moorwood96}, k. \cite{woo02}.}

\end{center}
\end{table*}

\section{X-ray bolometric corrections for CT AGN}
\label{sec_ctkbol}

\begin{figure}
\begin{center}
\includegraphics[width=90mm]{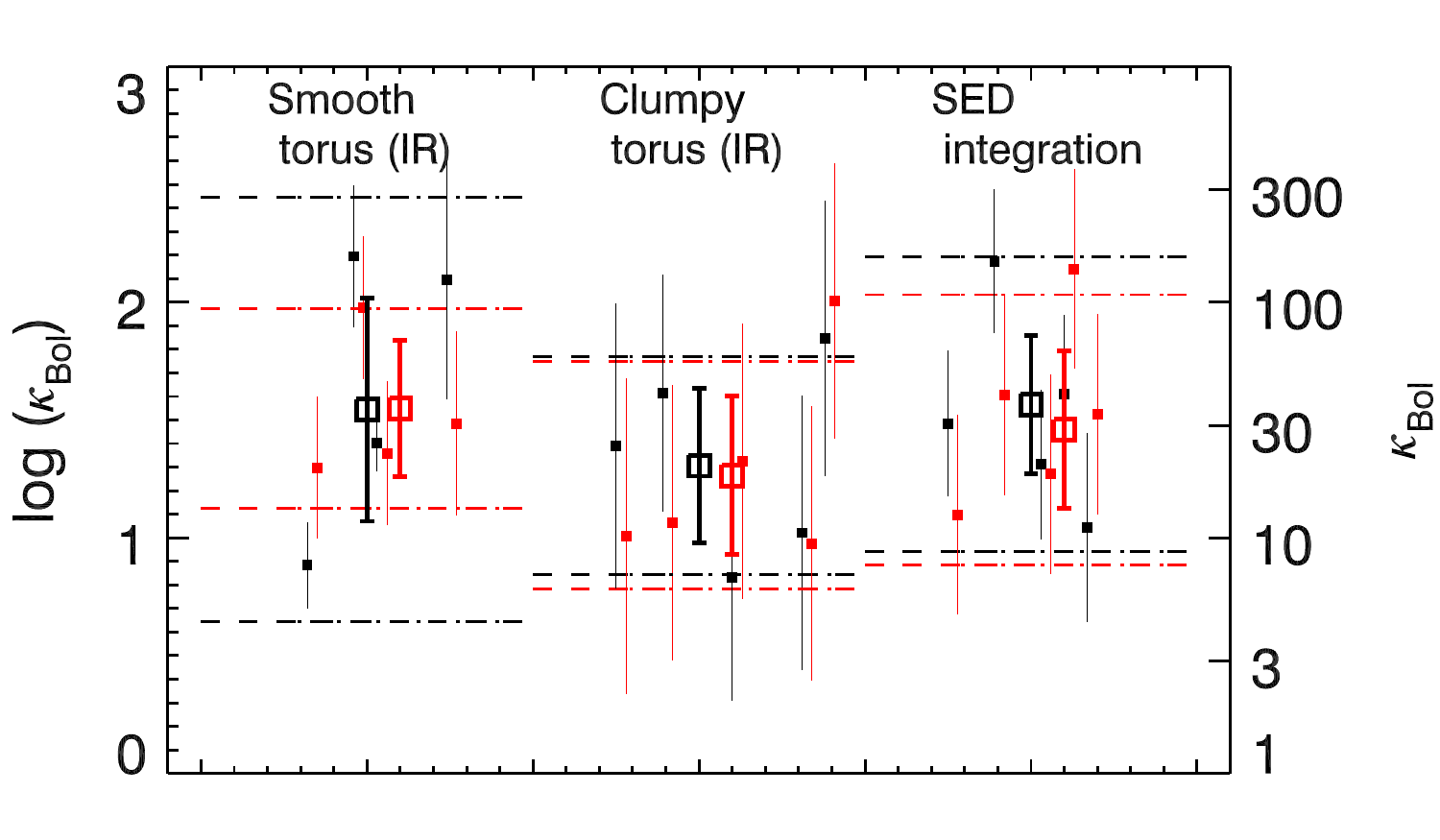}
\caption{Individual \kbol\ values (small squares ordered from left to right as they are ordered top to bottom in Table \ref{tab_sample}) calculated from the {\tt torus} (black) and {\tt mytorus} (red) models, given \lbol\ estimated from the smooth IR torus model (left), the clumpy IR torus model (middle) and from simple SED integration (right). The large empty squares show the mean of these values when taking into account intrinsic scatter, where the error bars represent the 1$\sigma$ uncertainty in the mean. The dotted lines show the estimated standard deviation of the intrinsic scatter.}
\label{fig_meankbol}
\end{center}
\end{figure}

With \lx\ and \lbol\ estimates from different methods for the 10 CT AGN, our first step is to investigate the \kbol\ values derived using each of these. Figure \ref{fig_meankbol} shows the individual \kbol\ values for each CT AGN and for each combination of \lx\ and \lbol. The uncertainties shown correspond to the uncertainties in \lx\ and \lbol\ combined in quadrature. Where no uncertainty is available, we assume a value of 0.3 dex which is typical of our sample. We also show the mean of each combination, calculated assuming that there is an intrinsic underlying Gaussian scatter in \kbol. The error bars represent the uncertainty in the mean. We also plot our estimate of the intrinsic scatter (1$\sigma$) in Figure \ref{fig_meankbol} which we find to be log\kbol$\sim0.5$ (with large uncertainties).

For the X-ray torus modeling, there is no evidence for a systematic difference in the mean \kbol\ values estimated from each model. This is true whether using the estimates of \lbol\ from \cite{ichikawa15}, \cite{gruppioni16} or \cite{woo02}. Furthermore, when considering source by source estimates, all \kbol\ estimates agree within the uncertainties when comparing the results from the X-ray torus models. We also find no evidence for a systematic difference in \kbol\ values between the different \lbol\ estimates. We find that all the \kbol\ values, regardless of which X-ray or IR torus modeling is used, even simple SED integration, are statistically consistent with each other.

\begin{figure}
\begin{center}
\includegraphics[width=90mm]{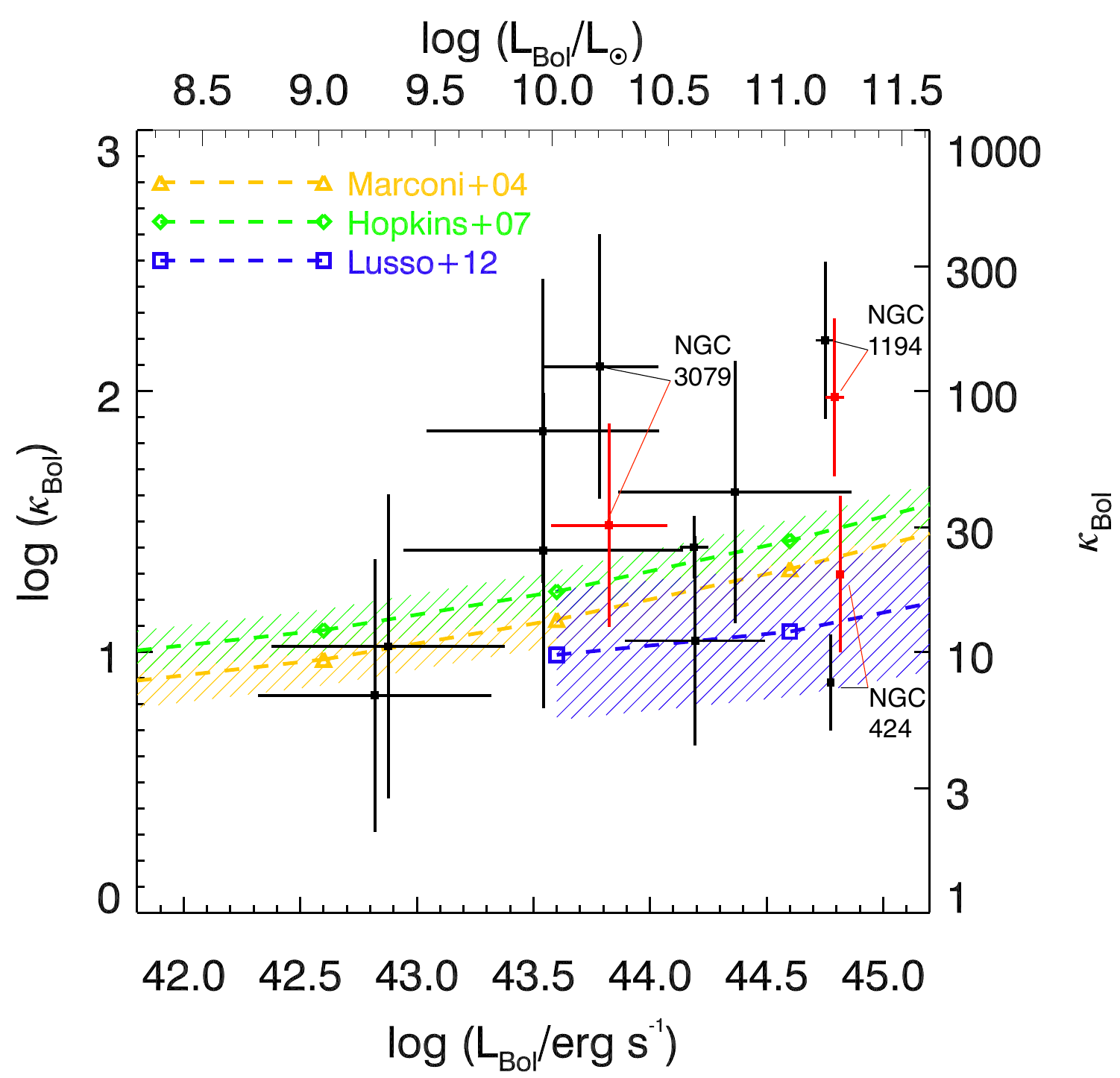}
\caption{X-ray bolometric corrections for the CT AGN versus \lbol, where \lx\ has been estimated from the {\tt torus} model (black points). We show our results with respect to the published relationships from \cite{marconi04}, \cite{hopkins07} and \cite{lusso12}. Dashed regions show their 1$\sigma$ intrinsic dispersions. For most sources the measurements agree with the relationships. However, for NGC~424 and NGC~3079, \kbol\ given the \lx\ estimate from the {\tt mytorus} model provides a better agreement, which we plot in red, shifted to slightly higher \lbol\ values for clarity. For NGC~1194, both estimates lie significantly away from the relationships.}
\label{fig_lbol}
\end{center}
\end{figure}

\begin{figure}
\begin{center}
\includegraphics[width=90mm]{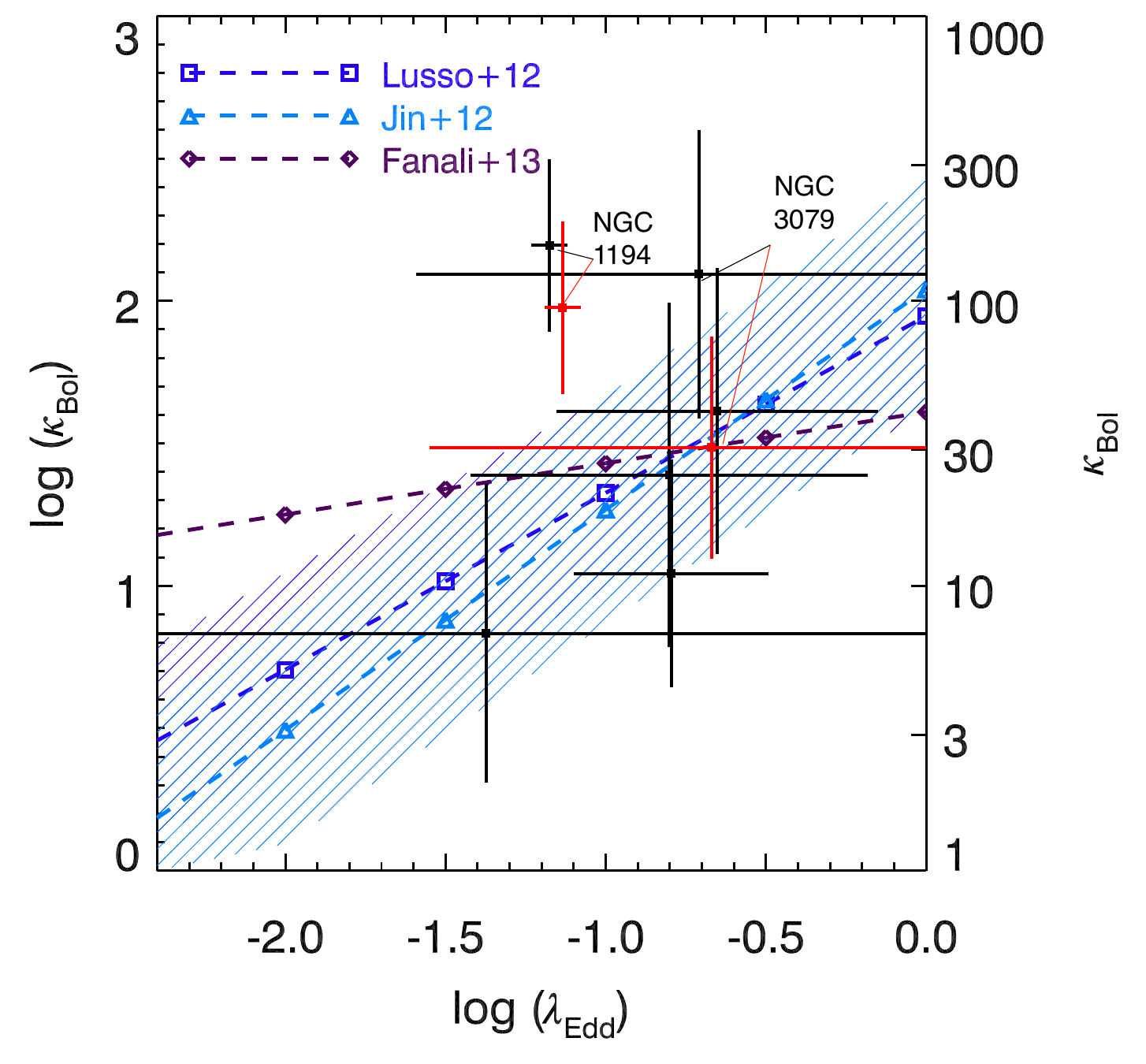}
\caption{X-ray bolometric corrections for the CT AGN versus \lamedd, where \lx\ has been estimated from the {\tt torus} model (black data points). We also plot the relationships from \cite{lusso12}, \cite{jin12} and \cite{fanali13}. Dashed regions show their 1$\sigma$ intrinsic dispersions (no measure of the dispersion is presented by \cite{fanali13}). Similarly for our comparison with relationships for \lbol, we find that most of our measurements agree for both X-ray models and that NGC~1194 lies significantly above the relationships.}
\label{fig_lamedd}
\end{center}
\end{figure}

Since there are well established relationships between \kbol\ and \lbol\ and \kbol\ and \lamedd, we proceed to test our derived \kbol\ values by comparing to these relationships. For this we investigate estimates of \kbol\ when using either \lx\ from the {\tt torus} model and \lx\ from the {\tt mytorus} model.  We plot our \kbol\ values against \lbol\ in Figure \ref{fig_lbol} along with the relationships presented in \cite{marconi04}, \cite{hopkins07} and \cite{lusso12} and their intrinsic dispersions. With regards to the dependence of \kbol\ on \lamedd, we plot our results with the previously reported relationships between these quantities from \cite{lusso12}, \cite{jin12} and \cite{fanali13} in Figure \ref{fig_lamedd}.

For most sources the measurements agree with the relationships for both \lx\ measurements. However, for NGC~424 and NGC~3079 the \kbol\ values given the \lx\ estimate from the {\tt mytorus} model provide better agreement. For NGC~1194, both estimates lie significantly away from the relationships, $\sim2\sigma$ for the {\tt torus} model and $\sim1.5\sigma$ for the {\tt mytorus} model. For our analysis of \kbol\ henceforth, we use the \lx\ estimate from the {\tt torus} model with the exception of NGC~424 and NGC~3079 where we use the \lx\ estimate from the {\tt mytorus} model. For NGC~1194 the fact that neither \lx\ estimates are in agreement with the relationships may imply that the intrinsic \lx\ has been underestimated by $\sim0.5$ dex. Alternatively \lbol\ may have been overestimated by the same amount. We discuss and investigate the inclusion of NGC~1194 in our sample in later analysis.

While many previous works have calculated \kbol\ for unobscured AGN and obscured but Compton-thin AGN, this is the first time systematic calculations of \kbol\  for CT AGN have been carried out. By combining our results with those for unobscured AGN and Compton-thin AGN, this allows us to investigate if \kbol\ is dependent on \nh, and over a wider range than was previously possible. In the context of the standard AGN unification model, whereby higher obscurations corresponds to larger viewing angles through the torus, probing the dependence of \kbol\ on \nh\ will allow us to test orientation effects. Specifically we will explore if the fraction of the accretion disk emission reprocessed by the corona into the X-rays is orientation dependent. 

For unobscured AGN and Compton-thin obscured AGN we again use the large sample of \lbol\ estimates from IR torus SED fitting presented in \cite{gruppioni16}, the parent sample of which was the extended 12 micron galaxy sample by \cite{rush93}. Absorption column measurements, \nh\ and intrinsic \lx\ values for a large subset of this sample was presented in \cite{brightman11} from X-ray spectral analysis of \xmm\ data. In order to do as direct a comparison as possible, we restrict our comparison to sources that have the same range in \lbol\ as our sources, i.e. \lbol$\approx10^{42}-10^{45}$ \ergs, a total of 21 sources. 

\begin{figure}
\begin{center}
\includegraphics[width=90mm]{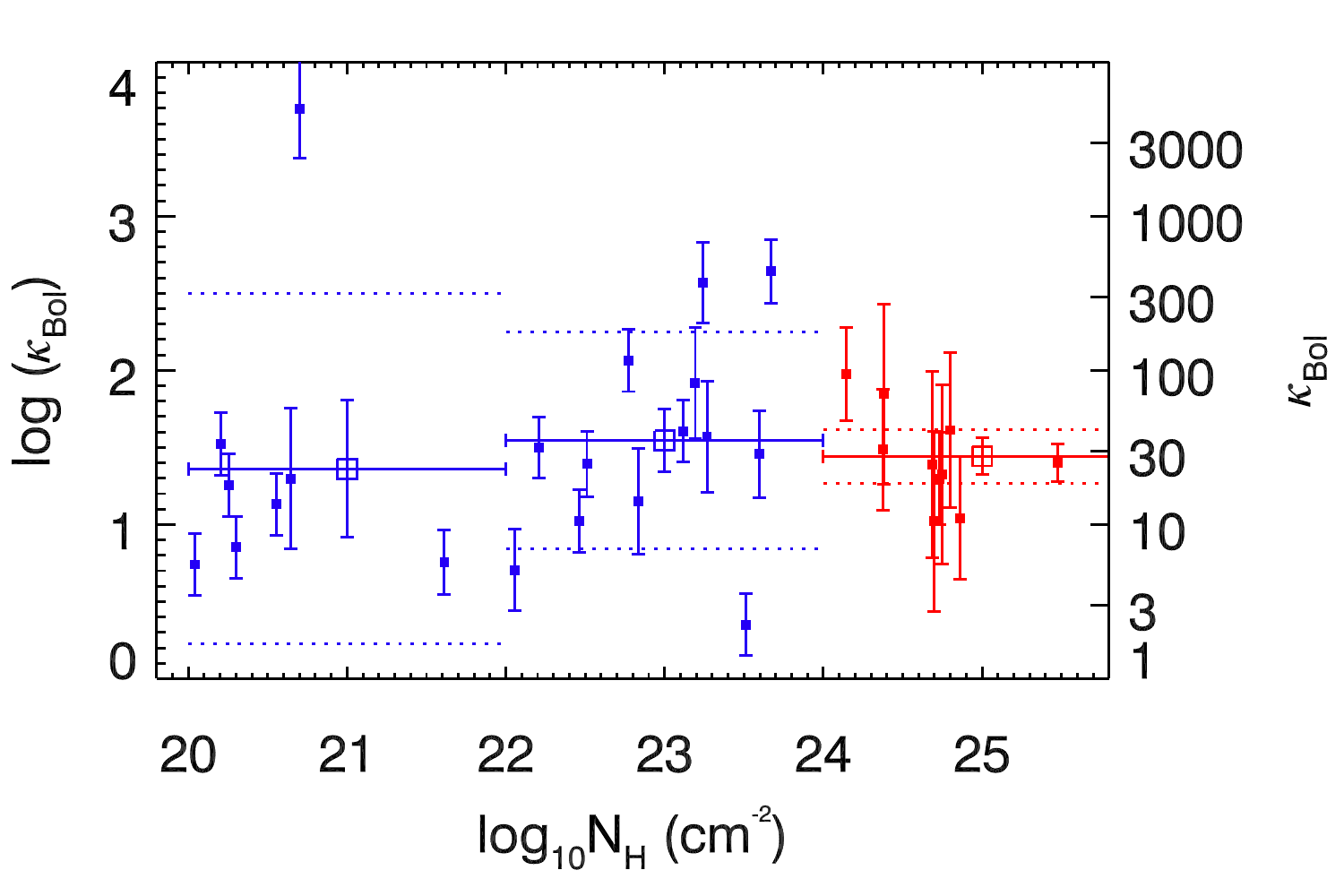}
\caption{Individual \kbol\ values as a function of \nh\ for our CT AGN sample (small filled red squares) combined with data from unobscured and Compton-thin AGN  from \cite{gruppioni16} and \cite{brightman11} (small filled blue squares) selected to have the same range of \lbol\ as our sample. The large empty squares represent the means of these data points for three bins in \nh, and the dotted lines mark the estimated $\sigma$ of the intrinsic scatter. The extreme outlier is NGC~6810 where it is likely \lbol\ from SED fitting has been severely overestimated.}
\label{fig_nh}
\end{center}
\end{figure}

We plot \kbol\ against \nh\ combining our results on CT AGN with the results from unobscured and Compton-thin AGN in Figure \ref{fig_nh}. In order to investigate the dependence of \kbol\ on \nh, we calculate mean \kbol\ values for 3 bins in \nh, log(\nh/\cmsq)=20--22, 22--24 and 24--26, and estimate the intrinsic scatter assuming it to be a Gaussian centered on the mean, finding that log$_{10}$\kbol$=1.36\pm0.44$, log$_{10}$\kbol$=1.54\pm0.20$ and log$_{10}$\kbol$=1.44\pm0.12$ respectively with an intrinsic scatter of $\sim0.2-1$ dex. The mean \kbol\ values are all within 1--2$\sigma$ of each other implying that there is no strong dependence of \kbol\ on \nh.

Among the unobscured AGN, NGC~6810 appears to be an extreme outlier with \kbol$>3000$. Here it is possible that \lbol\ estimated through SED fitting in \cite{gruppioni16} has been overestimated since these authors find that the \lbol\ estimate from the [Ne\,{\sc v}] and [O\,{\sc iv}] lines are more than a magnitude less than that from SED fitting. We therefore consider the effect of excluding this source from further analysis.

For us to put an estimate on the anisotropy of the corona we assume that our CT AGN are viewed edge-on and that sources with log(\nh/\cmsq)=20--22 are viewed face-on. Since many of our CT AGN are megamaser sources, which are required to be viewed at high inclination, our first assumption is well motivated. To assume that unobscured AGN are viewed face-on we must invoke the unification scheme. We then define anisotropy as the fraction of \lx\ emitted by unobscured AGN to that emitted by CT AGN given the same \lbol. This simply equates to 

\begin{equation}
anisotropy\equiv\frac{\kappa_{\rm Bol}(log_{10}(N_{\rm H}/{\rm cm}^{-2})=24-26)}{\kappa_{\rm Bol}(log_{10}(N_{\rm H}/{\rm cm}^{-2})=20-22)}
\end{equation}

Given our data we find this to be 1.2 (1$\sigma$ confidence range $=0.4-3.5$), suggesting that the corona emits $\sim$1.2 times more in polar directions with respect to equatorial directions with a 1$\sigma$ upper limit of 3.5 times. If we were to exclude the outliers NGC~1194 and NGC~6810 from our analysis of the anisotropy, we would find that the anisotropy is 2.1 (1.4--3.2). Our results imply that the X-ray corona is not strongly anisotropic.

\subsection{\kbol\ as a function of observed \lx\ and redshift}
\label{sec_kbolz}

The \kbol\ values that we have derived here for CT AGN can be used to estimate bolometric luminosities; however, this is only the case when a good estimate of the intrinsic \lx\ is available. This requires relatively good, high-energy X-ray data, for example from \nustar, in order to conduct X-ray torus modeling to account for the reprocessing effects of the Compton-thick obscuring medium. Such data will be available for a large number of local Seyfert 2s from modeling by Balokovi\'{c} et al. (in prep). 

The all-sky \swiftbat\ survey has become a popular resource for detecting and identifying CT AGN in the local universe. For example, \cite{ricci15} and \cite{akylas16} identify $\sim50$ CT AGN in the 70-month \swiftbat\ catalog \citep{baumgartner13}, also presenting intrinsic \lx\ values from torus modeling. For the seven sources in our sample that have been detected by \swiftbat, we determine that the mean \kbol\ given the intrinsic 14--195 keV \lx\ estimates from \cite{ricci15} is $1.12\pm0.17$ with an intrinsic scatter estimated to be $0.30\pm0.25$.

In addition, \cite{koss16b} presented a method for identifying local CT AGN in low-quality \swiftbat\ spectra. Since it is difficult to estimate intrinsic \lx\ for these sources, we explore \kbol\ for the {\it observed} 14--195 keV luminosity. We compile observed \lx(14--195 keV) values for the seven CT AGN in our sample that were detected by \swiftbat. We then calculate the bolometric correction factors for these {\it observed} luminosities using the \lbol\ values presented in Table \ref{tab_lum}, which we find to be $1.70\pm0.19$ with an intrinsic scatter estimated to be $0.36\pm0.21$.

While \swiftbat\ has detected and identified numerous CT AGN in the local universe, the high spatial resolution and sensitivity of \chandra, \xmm\ and \nustar\ are better suited for detecting these sources at higher redshift. For example \cite{brightman12b} and \cite{brightman14} have identified $\sim100$ CT AGN candidates up to $z\sim4$ in the deep \chandra\ observations of the CDFS, AEGIS-XD and C-COSMOS fields. However, due to the low-count nature of these sources, spectral parameters are difficult to constrain well, not least the intrinsic \lx. Intrinsic \lx\ estimates are usually obtained by fixing one or more spectral parameters, such as $\Gamma$ and the opening angle of the torus, $\theta_{\rm tor}$, to canonical values ( e.g. 1.9 for $\Gamma$ and 60\degree\ for $\theta_{\rm tor}$) . However, spectral analysis of CT AGN  with \nustar\ have revealed a wide variety of spectral shapes and complexity that is neglected when assuming a simple spectral model as described above.

We therefore use the best-fit models of our ten sources, which includes the range of spectral parameters observed and all spectral complexity such as a scattered power-law component, to calculate the {\it observed} \chandra\ luminosity that would be seen were they observed at higher redshifts. Our broadband \nustar\ spectra are essential for this since they tell us what \chandra\ is observing at these epochs. For example, at $z=2$, the observed 0.5--8 keV \chandra\ bandpass corresponds to rest-frame 1.5--24 keV, the expected flux in which is straightforward to calculate from our \nustar\ spectra.

\begin{figure}
\begin{center}
\includegraphics[width=90mm]{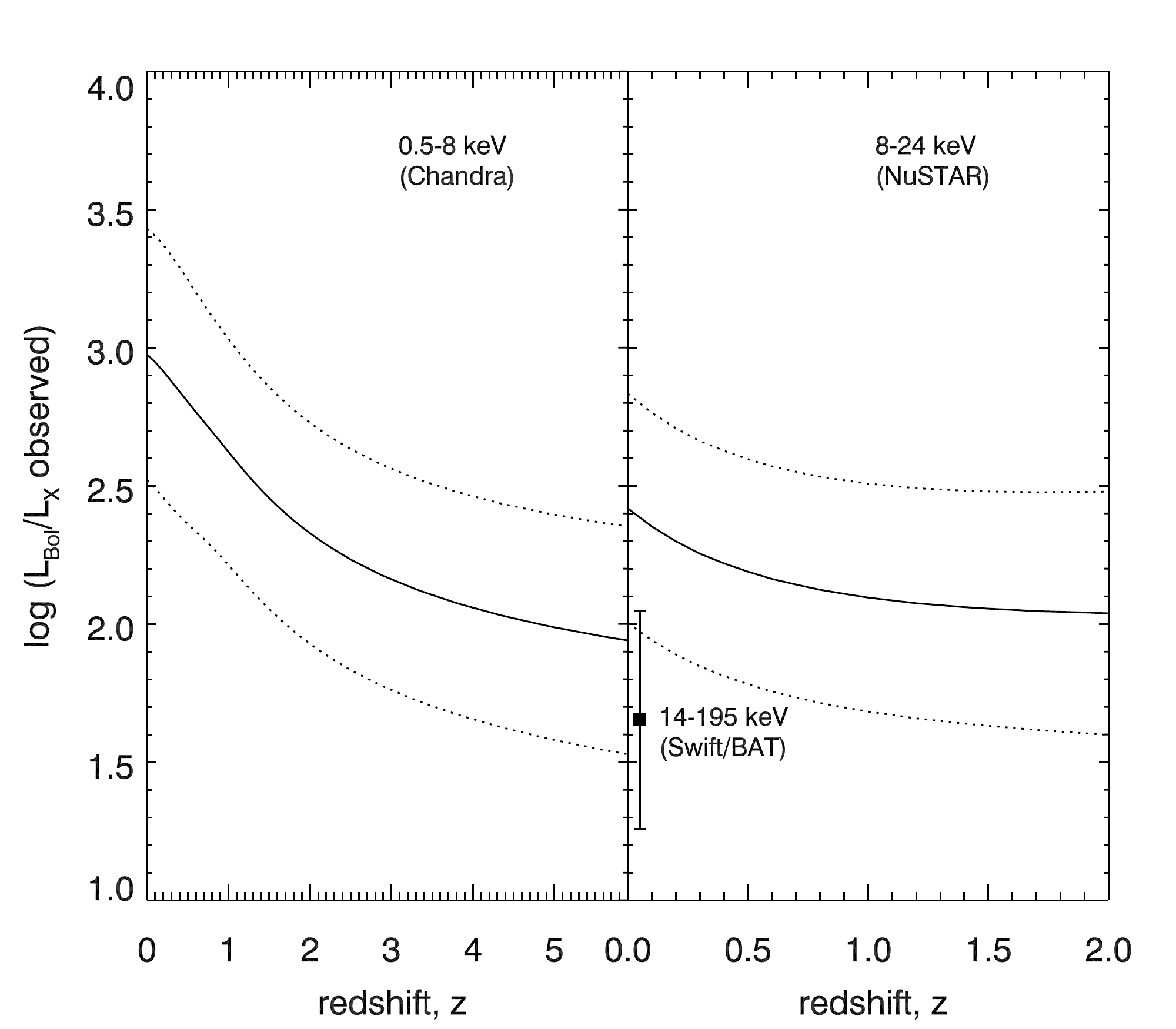}
\caption{Mean bolometric corrections (solid lines) with 1$\sigma$ spread (dotted lines) for the CT AGN in our sample given the observed \chandra\ 0.5--8 keV (left) and \nustar\ 8--24 keV (right) luminosities as a function of redshift. We also show the \swiftbat\ \kbol\ value for our sample at low redshift.}
\label{fig_kbolz}
\end{center}
\end{figure}

We then define a bolometric correction to this {\it observed} luminosity, 

\begin{equation}
\kappa_{\rm Bol}^{\prime}\equiv\frac{L_{\rm Bol}}{L(0.5-8\, {\rm keV}, observed)}
\end{equation}
and calculate this for each source from its X-ray spectrum and known \lbol\ for a range of redshifts. We include in the \lbol\ value the intrinsic \lx, despite the fact that the bolometric correction is to the observed \lx. We then calculate the mean of this $\kappa_{\rm Bol}^{\prime}$ from all ten sources at each redshift. Figure \ref{fig_kbolz} shows this mean $\kappa_{\rm Bol}^{\prime}$ and its corresponding 1$\sigma$ spread for redshifts up to $z=6$. Table \ref{tab_kbolz} gives these numbers for ease of interpretation.

A small number of CT AGN have also been identified in the \nustar\ surveys of the same fields above \citep[e.g.][ Del Moro et al. in prep, Zapacosta et al. in prep]{civano15} and as such we also carry out the same calculations as above, but for the observed 8--24 keV \lx, and also present these values in Figure \ref{fig_kbolz}. Since the restframe 8--24 keV band can only be observed with \nustar\ up to $z=2$ (restframe 24 keV corresponds to observed 72 keV, which is at the end of the \nustar\ bandpass), we only show up to this redshift. 

The main caveat involved with this method is that our sample contains relatively low luminosity AGN. Since the typical luminosities of CT AGN detected and identified at high-redshift are $\approx1-2$ orders of magnitude more luminous than ours, luminosity effects must be taken into account. Firstly, the distribution of spectral parameters are expected to be different at higher luminosities, for example \thetator\ is expected to be larger \citep{brightman15}. This is a relatively small effect, however, and not larger than the 1$\sigma$ range of values presented in Figure \ref{fig_kbolz}. Secondly the known relationship between \kbol\ and \lbol\ (Figure \ref{fig_lbol}) means that \kbol\ is systematically higher for these more luminous AGN. The median \lbol\ of our sample is $\sim10^{44}$ \ergs\ ($\sim10^{10.5}$ \lsol). For the most luminous AGN (e.g. \lbol$\gtrsim10^{46}$ \ergs) \kbol\ is a factor of $\approx6$ greater than at the luminosities of our sample, which should be taken in to account. Since the dependence of \kbol\ on \lbol\ is well known, it can be used to correct the estimated \lbol. 

For example, if we were to consider a source at $z=2$ with an observed 0.5--8 keV luminosity of $10^{44}$ \ergs, from Figure \ref{fig_kbolz} we would estimate its \lbol\ as 2.5$\times10^{46}$~\ergs\ (i.e. log($\kappa_{\rm Bol}^{\prime})\approx2.4$). For this \lbol\ value, the relationship presented by \cite{marconi04} would predict \kbol$\approx60$. Since the mean \kbol\ of our lower luminosity sample is $\approx25$, the original \lbol\ estimate of 2.5$\times10^{46}$~\ergs\ should be corrected upwards by a factor of $\frac{60}{25}=2.4$, making \lbol$\sim6\times10^{46}$~\ergs.

\begin{table}
\centering
\caption{Mean bolometric corrections as a function of redshift}
\label{tab_kbolz}
\begin{center}
\begin{tabular}{c c c}
\hline
Redshift & log$_{10}L_{\rm Bol}/L$(0.5--8 {\rm keV}) & log$_{10}L_{\rm Bol}/L$(8--24 {\rm keV}) \\
\hline
   0.0& 2.976$\pm$ 0.454& 2.418$\pm$ 0.415\\
   0.1& 2.948$\pm$ 0.457& 2.353$\pm$ 0.412\\
   0.2& 2.914$\pm$ 0.457& 2.299$\pm$ 0.409\\
   0.3& 2.878$\pm$ 0.455& 2.254$\pm$ 0.408\\
   0.4& 2.841$\pm$ 0.450& 2.219$\pm$ 0.407\\
   0.5& 2.804$\pm$ 0.442& 2.189$\pm$ 0.407\\
   0.6& 2.766$\pm$ 0.432& 2.163$\pm$ 0.407\\
   0.7& 2.731$\pm$ 0.424& 2.143$\pm$ 0.408\\
   0.8& 2.695$\pm$ 0.417& 2.124$\pm$ 0.409\\
   0.9& 2.660$\pm$ 0.413& 2.109$\pm$ 0.411\\
   1.0& 2.622$\pm$ 0.409& 2.095$\pm$ 0.412\\
   1.1& 2.587$\pm$ 0.407& 2.085$\pm$ 0.414\\
   1.2& 2.551$\pm$ 0.405& 2.075$\pm$ 0.416\\
   1.3& 2.518$\pm$ 0.404& 2.068$\pm$ 0.419\\
   1.4& 2.486$\pm$ 0.402& 2.061$\pm$ 0.421\\
   1.5& 2.456$\pm$ 0.402& 2.056$\pm$ 0.424\\
   1.6& 2.427$\pm$ 0.401& 2.051$\pm$ 0.427\\
   1.7& 2.401$\pm$ 0.401& 2.047$\pm$ 0.430\\
   1.8& 2.375$\pm$ 0.400& 2.044$\pm$ 0.433\\
   1.9& 2.351$\pm$ 0.400& 2.042$\pm$ 0.437\\
   2.0& 2.329$\pm$ 0.400& 2.039$\pm$ 0.440\\
   2.1& 2.307$\pm$ 0.400& 2.038$\pm$ 0.444\\
   2.2& 2.287$\pm$ 0.400& 2.037$\pm$ 0.448\\
   2.3& 2.269$\pm$ 0.400& 2.035$\pm$ 0.451\\
   2.4& 2.251$\pm$ 0.400& 2.034$\pm$ 0.455\\
   2.5& 2.233$\pm$ 0.400& 2.034$\pm$ 0.459\\
   2.6& 2.218$\pm$ 0.400& 2.033$\pm$ 0.463\\
   2.7& 2.203$\pm$ 0.400& 2.033$\pm$ 0.467\\
   2.8& 2.189$\pm$ 0.400& 2.033$\pm$ 0.471\\
   2.9& 2.174$\pm$ 0.400& 2.034$\pm$ 0.475\\
   3.0& 2.162$\pm$ 0.400& 2.034$\pm$ 0.479\\
   3.1& 2.150$\pm$ 0.401& 2.033$\pm$ 0.483\\
   3.2& 2.138$\pm$ 0.401& 2.036$\pm$ 0.486\\
   3.3& 2.126$\pm$ 0.401& 2.047$\pm$ 0.488\\
   3.4& 2.115$\pm$ 0.401& 2.056$\pm$ 0.490\\
   3.5& 2.105$\pm$ 0.402& 2.065$\pm$ 0.492\\
   3.6& 2.095$\pm$ 0.402& 2.075$\pm$ 0.494\\
   3.7& 2.086$\pm$ 0.402& 2.085$\pm$ 0.496\\
   3.8& 2.076$\pm$ 0.403& 2.096$\pm$ 0.498\\
   3.9& 2.067$\pm$ 0.403& 2.105$\pm$ 0.499\\
   4.0& 2.059$\pm$ 0.403& 2.115$\pm$ 0.501\\
   4.1& 2.051$\pm$ 0.404& 2.125$\pm$ 0.503\\
   4.2& 2.043$\pm$ 0.404& 2.135$\pm$ 0.505\\
   4.3& 2.035$\pm$ 0.405& 2.145$\pm$ 0.507\\
   4.4& 2.027$\pm$ 0.405& 2.156$\pm$ 0.508\\
   4.5& 2.021$\pm$ 0.405& 2.165$\pm$ 0.510\\
   4.6& 2.014$\pm$ 0.406& 2.175$\pm$ 0.511\\
   4.7& 2.007$\pm$ 0.406& 2.185$\pm$ 0.513\\
   4.8& 2.001$\pm$ 0.407& 2.195$\pm$ 0.515\\
   4.9& 1.994$\pm$ 0.407& 2.206$\pm$ 0.517\\
   5.0& 1.988$\pm$ 0.408& 2.217$\pm$ 0.519\\
   5.1& 1.983$\pm$ 0.408& 2.226$\pm$ 0.520\\
   5.2& 1.977$\pm$ 0.409& 2.236$\pm$ 0.522\\
   5.3& 1.971$\pm$ 0.409& 2.246$\pm$ 0.523\\
   5.4& 1.966$\pm$ 0.410& 2.257$\pm$ 0.525\\
   5.5& 1.960$\pm$ 0.410& 2.268$\pm$ 0.527\\
   5.6& 1.955$\pm$ 0.411& 2.280$\pm$ 0.529\\
   5.7& 1.950$\pm$ 0.411& 2.290$\pm$ 0.530\\
   5.8& 1.945$\pm$ 0.412& 2.300$\pm$ 0.531\\
   5.9& 1.941$\pm$ 0.412& 2.311$\pm$ 0.533\\

\hline
\end{tabular}
\tablecomments{A tabulated version of Figure \ref{fig_kbolz} for easier interpretation.}

\end{center}
\end{table}

\section{Discussion}
\label{sec_disc}

In our calculation of \kbol\ for CT AGN we have investigated different methods for estimating both \lx\ and \lbol\ for these heavily obscured sources, finding that the results are generally insensitive to the toroidal geometry assumed for the obscurer in both the infrared and X-rays. We also used established relationships between \kbol\ and \lbol\ and \kbol\ and \lamedd\ to test our derived \kbol\ values finding that they agreed well, implying that the torus modeling recovers these intrinsic parameters well. This is significant considering that the geometries assumed by the models differ, which is especially the case between the X-ray and IR models. Regarding a comparison of the torus models in the infrared, \cite{feltre12} conducted a comparison of the \cite{fritz06} and \cite{nenkova08} IR torus models which were used to obtain our \lbol\ estimates. These two models assume different dust distributions, smooth and clumpy respectively. \cite{feltre12} found that while the two models can produce similarly shaped SEDs, the underlying parameters derived, such as the covering factor, are different. However, in terms of the \lbol\ values derived from these models, we do not find a statistically significant difference between the models. 

Nevertheless, a few exceptions to this were found. We found that for NGC~424 and NGC~3079, the \lx\ estimate from the {\tt mytorus} model gave a \kbol\ value that is in better agreement with the relationships. For NGC~1194, our \kbol\ estimates lie significantly above the relationships by $\sim0.5$ dex. This could be due to a systematic underestimation of the intrinsic \lx, possibly caused by the underestimation of \nh. Alternatively, this could have been caused by an overestimation of \lbol\ in the SED fitting by \cite{gruppioni16}, perhaps due to contamination by star formation in the host galaxy. Finally, it is possible that the \kbol\ value for NGC~1194 lies at the extreme of the intrinsic ditribution of \kbol\ for its luminosity. Figure \ref{fig_nh} shows that similarly high \kbol\ values are found for the less obscured sources too.

We note that there are differences in the relationships between \kbol\ and \lbol\ presented by \cite{marconi04}, \cite{hopkins07} and \cite{lusso12}, some of which are to do with the definition of \lbol. \cite{marconi04} define their intrinsic bolometric luminosities as the sum of the optical and UV emission from the accretion disk and X-ray emission from the corona. \cite{hopkins07} follow a similar approach to \cite{marconi04}; however, they count the IR emission that is reprocessed disk emission. For this reason the \cite{hopkins07} relation is systematically higher than the \cite{marconi04} one. \cite{lusso12} use the sum of the AGN IR (1--1000 $\mu$m) and X-ray (0.5--100 keV) luminosities as a proxy for the intrinsic nuclear luminosity. Since they only count the reprocessed emission, their \kbol\ estimates should be comparable to \cite{marconi04}. However, it is lower. \cite{lusso12} discuss this finding, suggesting that since their sample is X-ray selected, it is biased towards X-ray bright sources that have lower \kbol\ values. We note that the differences in the established relationships are smaller than our uncertainties, so we cannot say which relationships our data agree with better. Regarding our methods for the CT AGN and their less obscured counterparts, we follow the same approach as \cite{marconi04}, in that we take \lbol\ to be the sum of the inferred optical and UV emission (from IR torus modeling) and X-ray emission.

We have also found that the intrinsic \kbol\ values for CT AGN is statistically consistent with \kbol\ for less obscured AGN indicating that there is little dependence of \kbol\ on \nh. Under the assumption of the standard AGN unification model, whereby for unobscured sources the central engine is viewed face-on and for heavily obscured sources it is viewed edge-on, this then implies that the fraction of X-rays emitted with respect to the optical/UV emission from the disk does not have a strong dependence on the orientation of the X-ray emitting corona. Since our sample contains many megamasers which are known to be viewed edge-on this supports our assumption based on unification. The lack of a strong dependence on orientation is important for understanding the physics of the disk-corona system, since it implies the corona emits almost isotropically, while the disk is known to emit anisotropically \citep{netzer87}. The models of \cite{you12} and \cite{xu15} predict a weak dependence of the optical to X-ray slope, $\alpha_{\rm OX}$ (which is strongly correlated with \kbol) on orientation. Since the predicted difference appears to be $<0.1$ dex in \kbol, and we place a 1$\sigma$ upper limit of 3.5 on this difference, the predictions of the models are not possible to detect with our current data.

Anisotropic X-ray emission would have possible implications for the AGN obscured fraction. \cite{sazonov15} proposed that collimation of X-rays in the polar direction (i.e. that observed in unobscured type 1 AGN) could lead to the observed dependence of the obscured fraction on \lx, and that potentially the intrinsic obscured fraction has no luminosity dependence as observed. This, however, would require a strong dependence of \lx\ on viewing angle, $\alpha$, following the cosine law, i.e. $dL/d\Omega\propto{\rm cos}\alpha$, such that \lx\ drops to zero for edge-on viewing angles. While our results allow for a factor of 3.5 drop from face-on to edge-on, they are inconsistent with the cosine law, albeit with a small sample. Similarly, \cite{brightman16b} found that megamaser CT AGN show the same relationship between the X-ray spectral index, $\Gamma$ and \lamedd\ as do unobscured AGN, further arguing against anisotropic X-ray emission.

Isotropic X-ray radiation, on the other hand, is also supported by the observed tight correlation between the X-ray and infrared luminosities that is statistically the same for both type 1 and type 2 AGN \citep[e.g.][]{gandhi09,asmus15}, unless both the IR and X-rays emit anisotropically in the same direction \citep{yangwangliu15}.

While we have found that the \kbol\ values for our sample of relatively low luminosity CT AGN are consistent with the relationship found for unobscured AGN in the same luminosity range, our sample lacks the high luminosity sources required to confirm if the increasing trend of \kbol\ with \lbol\ holds for CT AGN. One such highly luminous (\lbol$\sim10^{47}$ \ergs) close to Compton thick (\nh$\sim5\times10^{23}$ \cmsq) source, IRAS~09104+4109, where similar X-ray and IR torus modeling has been carried out, exists \citep{farrah16}. These authors estimate \lx\ to be $1-2\times10^{45}$ \ergs\ and \lbol\ to be $\sim1.8\times10^{47}$ \ergs\ implying that \kbol$\sim100$. This value agrees very well with the relationship found for unobscured AGN suggesting that there is agreement between heavily obscured AGN and unobscured AGN across a wide range in luminosities. 

 \section{Summary and Conclusions}
\label{sec_conc}

We have compiled intrinsic \lx\ and \lbol\ values for a sample of 10 local CT AGN from IR and X-ray torus modeling and have investigated \kbol\ for these heavily obscured sources for the first time. We find that:

\begin{itemize}
\item There are no statistically significant differences in \kbol\ values when using the X-ray torus models of \cite{murphy09} or \cite{brightman11} to calculate \lx\ or the infrared torus models of \cite{fritz06} or \cite{nenkova08} to calculate \lbol.

\item Our \kbol\ estimates for CT AGN are consistent with the established relationships between \kbol\ and \lbol\ in the range \lbol$\approx10^{42}-10^{45}$ \ergs\ and \kbol\ and \lamedd\ in the range \lamedd$\approx0.01-0.3$. However, we find that for  NGC~424 and NGC~3079 the \lx\ estimates from the {\tt mytorus} model provides better agreement. For NGC~1194 our \kbol\ estimate is too high considering both the \lbol\ or \lamedd\ relationships. This may imply that the intrinsic \lx\ has been underestimated by $\sim0.5$ dex or that \lbol\ has been overestimated by the same amount.

\item There is no evidence that \kbol\ depends on \nh. Under the assumptions of AGN unification, whereby the most obscured AGN are viewed edge-on and unobscured AGN are viewed face-on, this implies that the X-ray emission from the corona does not depend strongly on viewing angle. We estimate an upper limit on the anisotropy of the corona, finding that it emits no more than 3.5 times  (1$\sigma$ confidence level) in polar directions than in equatorial directions, albeit based on a small sample.

\item We have presented \kbol\ for CT AGN as a function of the observed \lx\ and redshift, useful for estimating \lbol\ of CT AGN identified in X-ray surveys where a good measurement of the intrinsic \lx\ is not available.

\end{itemize}

\acknowledgments

We thank the referee for providing a constructive review of our manuscript which improved its quality. We also thank Kohei Ichikawa for insights into IR torus modeling and Almudena Alonso-Herrero and Pat Roche for providing the IR data for Figure 1. M.\,Balokovi\'{c} acknowledges support from NASA Headquarters under the NASA Earth and Space Science Fellowship Program, grant NNX14AQ07H. We acknowledge financial support from the ASI/INAF grant I/037/12/0--011/13 (AC, AM) and the Caltech Kingsley visitor program (AC). This work was supported under NASA Contract No. NNG08FD60C, and made use of data from the {\it NuSTAR} mission, a project led by the California Institute of Technology, managed by the Jet Propulsion Laboratory, and funded by the National Aeronautics and Space Administration. We thank the {\it NuSTAR} Operations, Software and Calibration teams for support with the execution and analysis of these observations. Furthermore, this research has made use of the NASA/IPAC Extragalactic Database (NED) which is operated by the Jet Propulsion Laboratory, California Institute of Technology, under contract with the National Aeronautics and Space Administration. 

{\it Facilities:} \facility{\nustar}

\bibliography{bibdesk.bib}

\end{document}